\documentclass[11pt]{article}
\usepackage{moriond,epsfig}

\def\Journal#1#2#3#4{{#1} {\bf #2}, #3 (#4)}

\def\PLB{{\em Phys. Lett.}  B}
\def\PRL{\em Phys. Rev. Lett.}
\def\PRD{{\em Phys. Rev.} D}

\def\be{\begin{equation}}
\def\ee{\end{equation}}
\def\bea{\begin{eqnarray}}
\def\eea{\end{eqnarray}}

\newcommand{\ttbar}  {$t\bar{t}$}
\newcommand{\dzero}  {D\O}
\newcommand{\gev}    {$\rm{GeV/c^2}$}

\begin{document}
\vspace*{4cm}
\title{Top Quark Properties from the Tevatron}

\author{ M. Klute }

\address{MIT, Lab for Nuclear Science, 77 Massachusetts Ave,\\ Cambridge, MA 02139-4307, USA}

\maketitle\abstracts{
This report describes latest measurements and studies of top quark properties from the Tevatron in Run~II with an integrated luminosity of up to 750~pb$^{-1}$. Due to its large mass of about 172~\gev, the top quark provides a unique environment for tests of the Standard Model and is believed to yield sensitivity to new physics beyond the Standard Model. With data samples of close to 1~fb$^{-1}$ the CDF and \dzero~collaborations at the Tevatron enter a new aera of precision top quark measurements. 
}

\section{Introduction}
The top quark discovery in 1995 by the CDF~\cite{cite:discoverycdf} and \dzero~\cite{cite:discoveryd0}~experiments defines the start of the exciting era of top quark physics at the Tevatron. After very successful upgrades of the collider and both experiments data taking resumed in the year 2001. Since then, the Tevatron provided more than 1~fb$^{-1}$ of $p\bar{p}$ collision data at $\sqrt{s} = 1.96$~TeV to each experiment. With this data set the experiments enter a new aera of precision top quark measurements. 

Top property measurements to date rely on the strong production of top pairs. In the Standard Model (SM), the top quark is expected to decay to a W boson and a b quark nearly $100\%$ of the time. The W boson subsequently decays either to a pair of quarks or a lepton-neutrino pair. Depending on the lepton or hadronic decay of the two W bosons, the resulting event topologies of \ttbar~decays are classified as all-jets channel $(46.2\%)$, lepton+jets channel $(43.5\%)$ and di-lepton channel $(10.3\%)$. Each decay topology contains two b jets. Most top property measurements use the lepton+jets or di-lepton data sets. While the lepton in the above classification refers to $e, \mu$, or $\tau$, most of the results rely only on the $e$ and $\mu$ channels. 

\section{Top Quark Property Measurements}
In the following sections the latest measurements and studies of top quark properties are discussed. These measurements use data samples of up to 750~pb$^{-1}$. The questions, is the top quark consistent with SM predictions or does new physics manifest itself in the studied \ttbar~candidate sample is addressed.

\subsection{Resonant \ttbar~Production}\label{subsec:xttbar}
Resonances decaying in \ttbar~pairs are predicted by a number of theories beyond the SM. An example are top-color-assisted technicolor models~\cite{cite:zprime} which combine top-color and technicolor models. 
A massive \ttbar~condensate X or the heavy Z' boson which couples preferentially to the third generation is induced. The cross section for the Z' boson in this model is large enough for it to be observed over a wide range of masses and widths in the data available at the Tevatron. 

\begin{figure}
\begin{center} 
\psfig{figure=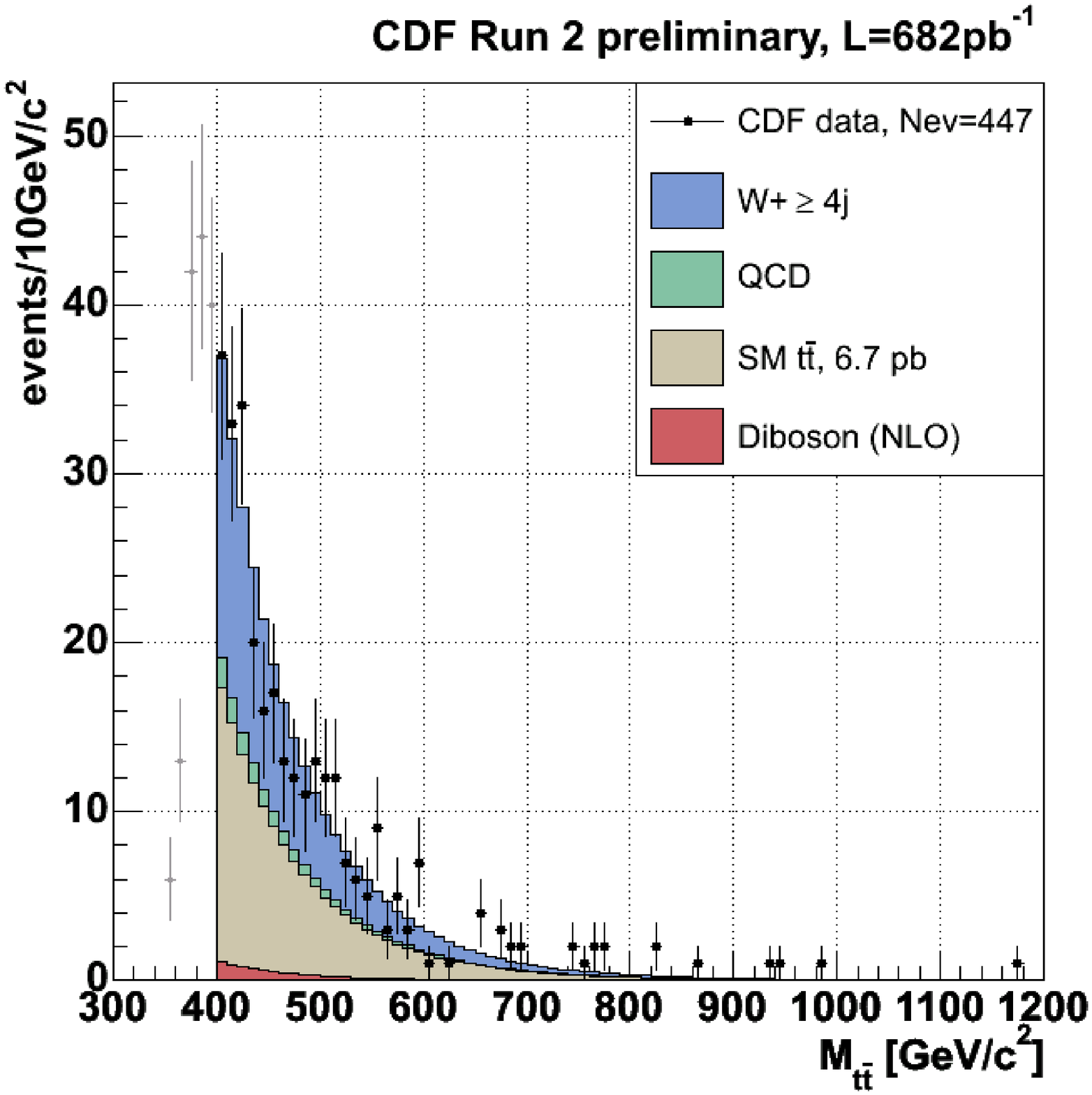,height=1.5in}
\psfig{figure=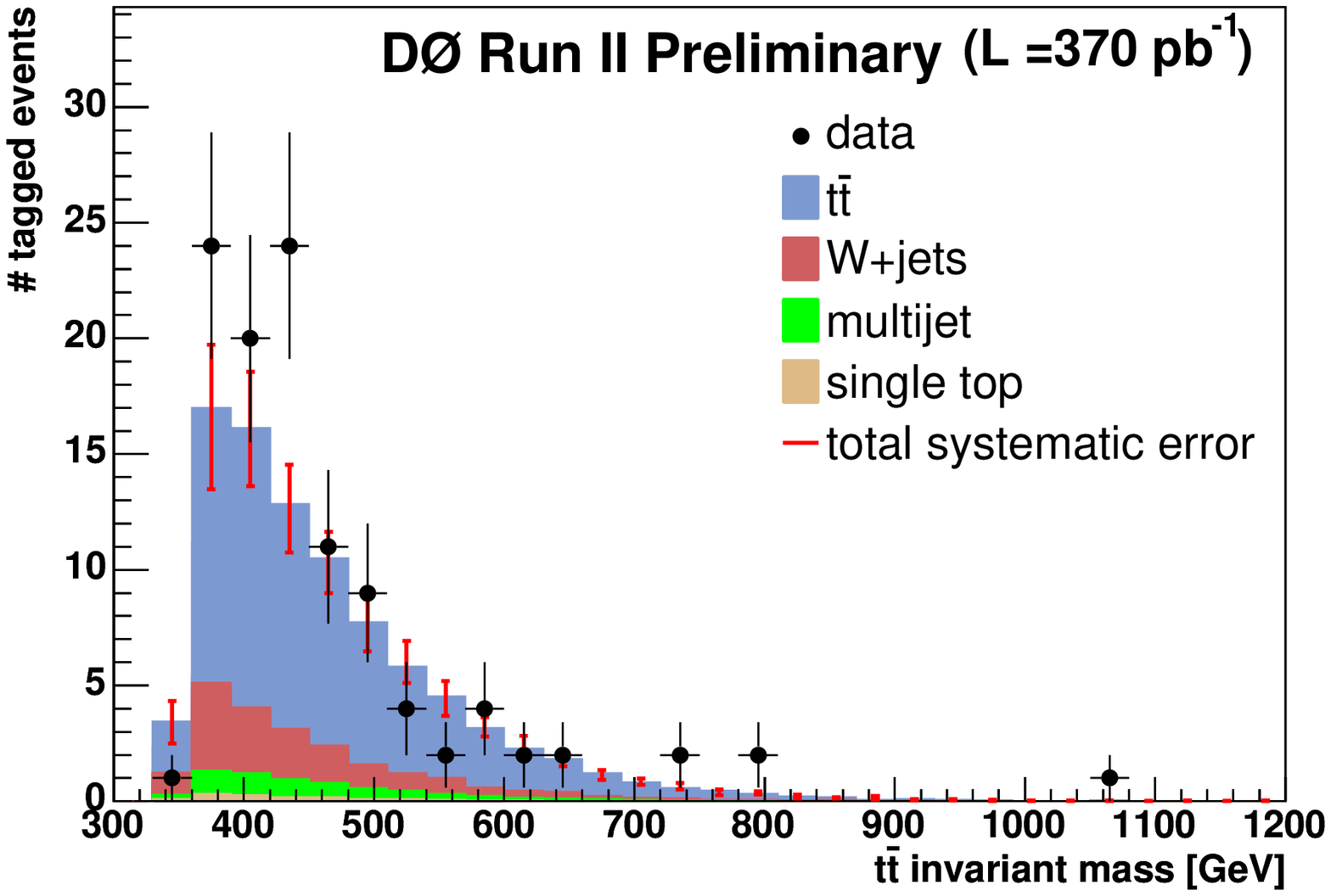,height=1.5in}
\caption{ Reconstructed \ttbar~invariant mass distribution by CDF (left) and \dzero~(right).
\label{fig:topttbar}}
\end{center}
\end{figure}

Both, CDF and \dzero~perform model-independent searches for a narrow-width resonance X decaying into a \ttbar~pair in the lepton+jets channel by examining the reconstructed \ttbar~invariant mass distribution. While \dzero~performs a counting experiment by assuming SM cross section for background processes, CDF searches for the new physics signal by performing a shape analysis. Figure~\ref{fig:topttbar} shows the reconstructed \ttbar~invariant mass distributions by CDF and \dzero. The results are consistent with the SM expectation and CDF and \dzero~exclude a signal with mass \linebreak $\rm{m_X} < 725$~\gev~\cite{cite:cdfttbar}~and $\rm{m_X} < 680$~\gev~\cite{cite:d0ttbar}, respectively. 

\subsection{Top Quark Lifetime}
Due to its large mass, the SM top quark decays with a lifetime of $\approx 10^{-24}$~s. 
The current experimental resolution does not allow a measurement of the top quark lifetime. Nevertheless, a study of the lifetime of top quarks can reveal new physics processes. 

CDF gives a first direct limit on the lifetime of the top quark.
  A sample of \ttbar~events is selected by identifying 
  electrons or muons and a secondary vertex (secvtx)
  tagged jet in events with $\ge$ 3 jets and large missing
  transverse energy.  In such events, the impact parameter
 (d$_0$) of the lepton is typically very
  small.  Its distribution can be predicted from the
  resolution of d$_0$ measurements for high momentum
  lepton tracks, the characteristics of SM top
  quark decays, and the magnitude and d$_0$
  distribution of non-\ttbar~background.
  Inconsistency between the predicted distribution and the
  observation could be indicative of a new, long-lived
  background to $t\bar{t}$, anomalous top production by a
  long-lived parent particle, or a large top quark lifetime.

  CDF shows that the lepton d$_0$ distribution in top candidate
  events from 318 pb$^{-1}$ of CDF data is consistent
  with expectations, and derives a limit on the top quark
  lifetime \linebreak 
  $c\tau < 52.5\,\mu$m at 95$\%$ and $c\tau < 43.5\,\mu$m at 90$\%$ confidence level~\cite{cite:lifetime}. 

\subsection{Searches for Heavy Quarks}
In the theoretical landscape, a number of models suggest the existence of a heavy $t'$ quark. Among these models are recent theoretical developments, such as Little Higgs Models, 2-Higgs Double scenarios, N=2 SUSY models, or the ``beautiful mirror model''~\cite{cite:bmirror}. Searches for t' are performed assuming that the new particle is pair produced, has mass greater than the top quark and decays promptly to $Wq$ final states, resulting in a topology very similar to \ttbar~events, but with larger values for the total transverse energy $H_T$.  

  CDF searches for a pair produced t'  decaying to Wq  in 347~pb$^{-1}$ of the CDF Run II data sample of lepton+jets events. The mass of the t' quark is reconstructed and a 2d-fit of the observed $H_T$ and reconstructed mass distribution is performed to discriminate the new physics signal from SM background. No evidence for a t' signal is observed.
 A t' with masses in the range from 196~\gev~to 207~\gev~can be excluded at 95\% confidence level~\cite{cite:tprime}, if the true \linebreak top mass is 175~\gev. 

\subsection{Top Quark Charge}
The electric charge is one of the most fundamental quantities characterizing a particle. The top quark is the only quark whose electric charge has not been measured so far. Models, alternative to the SM, predict an exotic quark, $Q_4$, with charge -4/3 with top like properties~\cite{cite:q4}. In order to determine the charge of the top quark one can study the charge of its decay products or investigate the photon radiation in \ttbar~events.

\begin{figure}
\begin{center} 
\psfig{figure=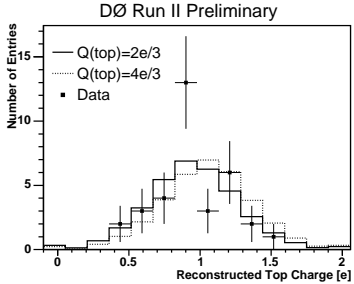,height=1.5in}
\caption{ The reconstructed top quark charge for the 34 top quark candidates.
\label{fig:topcharge}}
\end{center}
\end{figure}

\dzero~measures the charge of the top quark through its decay products using a data sample corresponding to an integrated luminosity of 360~pb$^{-1}$ in the lepton+jets channel. They select 17 events with at least four jets, two of which must be identified as coming from a b quark. A jet charge algorithm is applied to discriminate between $b$- and $\bar{b}$ jets. The performance of the algorithm is calibrated using $b\bar{b}$ pairs from data. A constrained kinematic fit is performed to reconstruct the \ttbar~event. \dzero~finds that the data is in good agreement with a top quark charge of 2e/3, as predicted by the SM and exclude the hypothesis of an exotic quark with charge of 4e/3 at 94\% confidence level~\cite{cite:charge}. Figure~\ref{fig:topcharge} shows the reconstructed top quark charge for \linebreak 34 top quark candidates.

\subsection{W Boson Helicity in Top Decays}
An important consequence of the large mass of the top quark is, to good approximation, that it decays as a free quark. The spin information carried by top quarks is expected to be passed directly on to their decay products, so that production and decay of top quarks provide a probe of the underlying dynamics, with minimal impact from gluon radiation and binding effects of QCD.
In the SM, the top quark has the same V-A charged-current weak interaction as all other fermions. This implies that the W boson in top decays cannot be right-handed, i.e. have positive helicity. The fraction of decays to longitudinal W bosons is determined by the masses of the involved particles and expected to be $f_0^{SM} = 70.3 \pm 1.2 \%$~\cite{cite:whel}.
Studies of decay angular and momentum distributions provide a direct check of the V-A nature of the Wtb coupling and information on the relative coupling of longitudinal and transverse W bosons to the top quark.

Both CDF and \dzero~measure the fraction $f_0$ of the longitudinal and $f_+$ of right-handed W bosons in the lepton+jets and di-lepton final state. The measurements are performed using a Matrix Element approach or by using the lepton $\rm{p_T}$, the invariant mass of the lepton and b-jet or $\cos\theta^{*}$, which results from the angle of the lepton in the rest frame of the W boson to the W boson flight direction in the top-quark rest frame, as discriminating variable.

At present, all studies of the helicity of the W boson in top quark decays are consistent with the SM and are limited by the statistic of the data samples. With the increasing data set the precision is improving rapidly. Eventually, when sufficient data is available, simultaneous measurements of $f_0$, $f_-$ and $f_+$ will be performed. At present, only measurements or limits from studies of single quantities are available. Table~\ref{tab:whel} summarizes these results.

\begin{table}[t]
\caption{Measurement and upper limits of the W helicity in top quark decays from CDF and \dzero. The integrated luminosity of the data set is given in units of pb$^{-1}$. \label{tab:whel}}
\vspace{0.4cm}
\begin{center}
\begin{tabular}{lccl}
\hline
source & data set & method & W helicity \\ \hline
CDF Run~I~\cite{cite:whelcdfrunIf0} & 106 & $p^l_T$ & f$_0 = 0.91 \pm 0.39$ \\
\dzero~Run~I~\cite{cite:wheld0runIf0} & 125 & ME & f$_0 = 0.56 \pm 0.32$ \\
CDF Run~II~\cite{cite:whelcdfrunIIf0} & 200 & $m^2_{lb}+p^l_T$ & f$_0 = 0.74^{+0.22}_{-0.34}$ \\ \hline

CDF Run~I~\cite{cite:whelcdfrunIfp} & 110 & $m^2_{lb}+p^l_T$ & f$_+ < 0.18$ \\
CDF Run~II~\cite{cite:whelcdfrunIIf0} & 200 & $m^2_{lb}+p^l_T$ & f$_+ < 0.27$ \\
\dzero~Run~II~\cite{cite:wheld0runIfp} & 230-370 & $\cos\theta^{*}+p^l_T$ & f$_+ < 0.25$ \\ \hline
\end{tabular}
\end{center}
\end{table}

\section{Conclusion}
After the top quark discovery in Run~I and the re-discovery of the top signal 
with the upgraded detectors and improved analysis techniques in the early Run~II, 
top quark physics at the Tevatron has now entered the stage of detailed studies 
of the top quark properties. To date all results are in agreement with the SM 
expectation and most measurements are limited by the statistic of the data samples. 
A large number of results on the top quark in the SM as well as searches for 
new physics within the sample of \ttbar~candidate events are becoming available. 
This development is expected to accelerate with $\ge$ 1~fb$^{-1}$ of data being 
available to both, CDF and \dzero, very soon.

\section*{Acknowledgments}
I thank the organizers of Rencontres de Moriond 2006 for a stimulating conference and gratefully acknowledge NSF award 
number PHY-0611671 for travel support. 
I thank the Fermilab staff and the technical staffs of the
participating institutions for their vital contributions. This
work was supported by the U.S. Department of Energy and National
Science Foundation; the Italian Istituto Nazionale di Fisica
Nucleare; the Ministry of Education, Culture, Sports, Science and
Technology of Japan; the Natural Sciences and Engineering Research
Council of Canada; the National Science Council of the Republic of
China; the Swiss National Science Foundation; the A.P. Sloan
Foundation; the Bundesministerium fuer Bildung und Forschung,
Germany; the Korean Science and Engineering Foundation and the
Korean Research Foundation; the Particle Physics and Astronomy
Research Council and the Royal Society, UK; the Russian Foundation
for Basic Research; the Comision Interministerial de Ciencia y
Tecnologia, Spain; and in part by the European Community's Human
Potential Programme under contract HPRN-CT-20002.

\end{document}